\documentclass[sigconf]{acmart}

\AtBeginDocument{%
  \providecommand\BibTeX{{%
    \normalfont B\kern-0.5em{\scshape i\kern-0.25em b}\kern-0.8em\TeX}}}

\setcopyright{rightsretained}

\begin{document}

\copyrightyear{2021}
\acmYear{2021}
\acmConference[FAccT '21]{Conference on Fairness, Accountability, and Transparency}{March 3--10, 2021}{Virtual Event, Canada}
\acmBooktitle{Conference on Fairness, Accountability, and Transparency (FAccT '21), March 3--10, 2021, Virtual Event, Canada}\acmDOI{10.1145/3442188.3445898}
\acmISBN{978-1-4503-8309-7/21/03}

\title{``This Whole Thing Smacks of Gender'': Algorithmic Exclusion in Bioimpedance-based Body Composition Analysis}

\author{Kendra Albert}
\authornote{Both authors contributed equally to this research.}
\email{kalbert@law.harvard.edu}
\affiliation{%
\institution{Harvard Law School}
\city{Cambridge, MA}}

\author{Maggie Delano}
\authornotemark[1]
\email{mdelano1@swarthmore.edu}
\affiliation{%
\institution{Swarthmore College}
\city{Swarthmore, PA}}
% }

\begin{abstract}
Smart weight scales offer bioimpedance-based body composition analysis as a supplement to pure body weight measurement. Companies such as Withings and Fitbit tout composition analysis as providing self-knowledge and the ability to make more informed decisions. However, these aspirational statements elide the reality that these numbers are a product of proprietary regression equations that require a binary sex/gender as their input. Our paper combines transgender studies-influenced personal narrative with an analysis of the scientific basis of bioimpedance technology used as part of the Withings smart scale. Attempting to include nonbinary people reveals that bioelectrical impedance analysis has always rested on physiologically shaky ground. White nonbinary people are merely the tip of the iceberg of those who may find that their smart scale is not so intelligent when it comes to their bodies. Using body composition analysis as an example, we explore how the problem of trans and nonbinary inclusion in personal health tech goes beyond the issues of adding a third ``gender'' box or slapping a rainbow flag on the packaging. We also provide recommendations as to how to approach creating more inclusive technologies even while still relying on exclusionary data.
\end{abstract}




\maketitle

%% TO DOS:
%% CCS codes??? - M
%% final read through, copy edits M/K
%% final article submission - M

\section{Introduction}

\textbf{Kendra:} \textit{As a nonbinary person who was assigned female at birth, I've always had an uncomfortable relationship with my body weight. To be honest, before I stepped on the Withings scale, I hadn't weighed myself in some time. But after undergoing some medical transition steps, I found myself much more curious about what my body fat percentage was. Especially since I have put on a fair amount of muscle, I was interested to see how the numbers matched or didn't match my self perception.}

\textit{Because of this discomfort with weight numbers, I asked Maggie to make a profile for me when I was getting started with the scale. She started going through the registration flow for a new user, and quickly encountered a roadblock - the system required a gender.}

\begin{figure}[tb]
  \centering
  \includegraphics[width=0.9\linewidth]{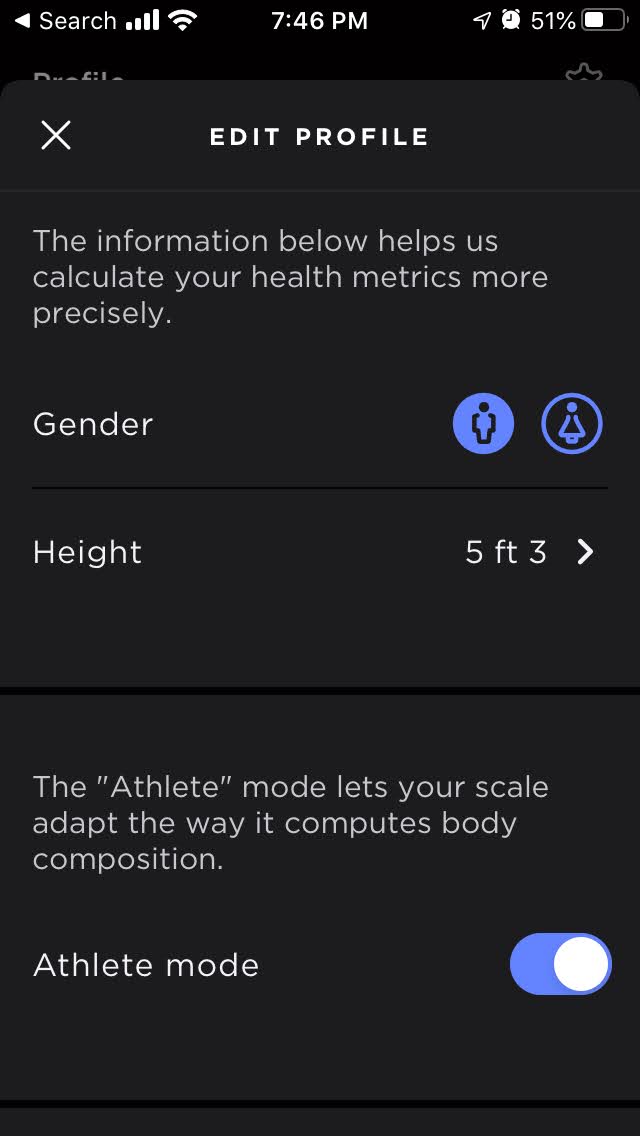}
  \caption{The Withings profile view in their Health Mate mobile application.}
  \Description{A view of the Withings profile view. There is an option for Gender with a \\
  person wearing pants and a person wearing a skirt, an option for height (5 ft 3) \\
  and an option for Athlete mode, which is set.}
  \label{fig:skirtOrPants}
\end{figure}

\textit{Wait, that’s not right. It didn't require a gender, it required you to pick one of two images labeled gender - a person wearing pants or a person wearing a skirt (see Figure \ref{fig:skirtOrPants}). When Maggie asked me which one I preferred, I told her to pick one and not tell me.}

\textit{When I did finally step on the scale (with my shiny new profile), I looked at the numbers and felt...well, some sort of way. To be honest, I’m happier with my body now than I remember ever being before, but the numbers seemed very off.} 

\textit{The next morning, we were talking about the scale at breakfast, and that was when Maggie first told me that the fat percentage was an estimation based on a technique called bioelectrical impedance analysis (BIA). There was an equation behind that number - it didn't actually measure my body fat percentage directly. I was shocked, and asked how that related to my gender. We decided to do some testing.}

\textit{When Maggie changed my “gender” from skirt-wearing to pants-wearing, my body fat percentage dropped 10\% points. 10! The huge difference seemed to confirm everything I’d thought about the numbers feeling wonky. But which one was right? Or failing that, which one closer? Could we look at how the algorithm calculated the percentage in order to see which one was likely to be more accurate for me?}

\textbf{Maggie:} \textit{Kendra’s reaction to the scale was an interesting experience because I realized that the knowledge I had about how BIA worked was non-obvious to anyone not familiar with the tech. I had already been thinking about how sex is often coded in the underlying BIA equations as 1 or 0 because I had given a guest lecture about it recently, but I hadn't fully thought through the implications for trans and nonbinary people.} 

\textit{Seeing the numbers on the scale jump by so much was jarring. We talked a lot about how to interpret the results and what the exact percentages really mean. There is a little progress bar on the scale that indicated that the numbers (regardless of gender) were above the “normal range.” But that normal range itself appears to be a “normal range” for an athlete because we were both well above that line. In terms of making the scale work for Kendra, the answer I was proposing was to pick skirt-wearing or pants-wearing and then track that number over time.} 

\textit{Kendra’s interest in using the Withings smart scale, and figuring out what “gender option” was right drove us to reexamine how BIA works as a technology, and how assumptions about gender and sex are built into the fundamental equations that drive the algorithmic models used for estimating body fat percentage.}

This paper, the result of that analysis, combines transgender studies-influenced personal narratives with an analysis of the scientific basis behind the bioimpedance technology used as part of the Withings smart scale. With these different forms of knowledge, we explore how the problem of trans and nonbinary inclusion in personal health tech goes beyond adding a third ``gender'' option or slapping a rainbow flag on some packaging. Rather, nonbinary exclusion is the tip of an iceberg of flawed assumptions and exclusionary clinical testing, resulting in algorithms that are advertised for self-knowledge but prove to allow anything but.

\section{Background}
\label{sec:background}

This paper draws on previous work related to self-tracking, transgender studies, human-computer interaction, and the study of sex and gender in biomedical research. In this section, we provide a brief summary of related work in these disciplines to situate our findings.

While the Withings weight scale is not the first commercially available scale to estimate body fat percentage, the device was one of the original ``smart'' (i.e. connected) weight scales \cite{BodyCompositionWiFi2020}. The device was first sold in the early 2010s at the beginning of a surge of interest in self-tracking and the advent of the ``quantified self'' movement \cite{lupton2016quantified,neff2016self,kragh2016quantified}. The quantified self movement included a variety of stakeholders including individual self-trackers, organizations such as Quantified Self, companies and toolmakers, academic researchers, and physicians (with considerable overlap between these categories) \cite{boeselWhatQuantifiedSelf2013}. Participants are broadly interested in the capabilities of self-tracking to provide unique, actionable, and personalized insights \cite{luptonDataMatteringSelftracking2020,wolfConceptualFrameworkPersonal2020}. Self-trackers engage deeply with their data through a process sociologist Deborah Lupton refers to as ``data sense-making'' \cite{luptonHowDataCome2018}. Many self-trackers believe that data can ``speak for itself'' and should be involved in medical care \cite{fiore-gartlandCommunicationMediationExpectations2015,omerEmpoweredCitizenHealth2016}. However, the use of data collected from commercial ``wellness'' devices such as the Withings scale or activity trackers like Fitbits is controversial as these devices don't always perform well in third-party validations, and often involve proprietary algorithms and metrics (e.g. steps, sleep scores).

Previous research investigating self-tracking devices and wearable monitors has shown that these devices, like devices in other categories, are designed primarily for an unmarked user \cite{ciforGenderedDesignDuoethnographic2020,costanza-chockDesignJusticeCommunityLed2020}. That is, the user is  assumed to be a White, cisgender, non-disabled man. Cifor and Garcia, for example, use duoethnography to evaluate gendered assumptions in the design and app of the Jawbone UP3 fitness tracker. They illustrate that while the device itself appeared ``genderless,'' the design of the device and the app reinforced masculinist values such as competition \cite{ciforGenderedDesignDuoethnographic2020}. Such issues are also present in the design of algorithms - for example, Fitbit devices count steps less reliably at slower gait speeds and with softer steps, which decreases step count accuracy for older people or people with mobility related disabilities \cite{feehanAccuracyFitbitDevices2018,javorinaInvestigatingValidityFitbit2020}. Early implementations of the hardware and algorithms used to estimate heart rate on wearables were less accurate for users with darker skin tones \cite{shcherbinaAccuracyWristWornSensorBased2017,hailuFitbitsOtherWearables2019}, though recent evidence suggests these disparities may have been addressed by improvements to the device's algorithms \cite{bentInvestigatingSourcesInaccuracy2020}. 

The development of algorithms without a diverse set of users creates \textbf{algorithmic exclusion}. Populations are excluded from the use of algorithmic tools because they were not included as part of the original data used in development, or because information was not gathered in such a way as to make their needs visible. This algorithmic exclusion means that the performance of these algorithms for individuals not in the original dataset are unknown; the practical implication is that these algorithms likely work less well for those not included in the original dataset. Algorithmic exclusion can have real world impacts as individuals rely more and more on these data, especially when these data are used by physicians. For example, pulse oximeter devices that measure blood oxygenation (using a more involved technique similar to that used by wearables manufacturers for measuring heart rate) overestimate blood oxygenation in individuals with darker skin tones \cite{bicklerEffectsSkinPigmentation2005,feinerDarkSkinDecreases2007}. Renewed interest in these disparities during the COVID-19 epidemic led to a study that showed that Black patients had nearly three times the frequency of occult hypoxemia (oxygen depravation) that was not detected by pulse oximetry than White patients \cite{moran-thomasHowPopularMedical2020,sjodingRacialBiasPulse2020}, potentially leading to higher mortality rates for Black patients when the seriousness of their COVID-19 cases were underestimated.

These issues have not escaped notice within communities that build technological tools. There has been increasing discussion in different design communities about how to create technology that is more inclusive and/or addresses some of the disparities discussed above. In human-computer interaction (HCI) and artificial intelligence (AI) research, for example, there have been efforts including analytical reviews, software analysis of datasets, and guidelines about increasing ``gender sensitivity'' \cite{burtscherWhereWouldEven2020, scheuermanHowComputersSee2019, scheuermanHowWeVe2020, hamidiGenderRecognitionGender2018, keyesMisgenderingMachinesTrans2018} and more intersectional approaches to addressing disparities, such as the intersection of gender and race \cite{buolamwiniGenderShadesIntersectional2018,10.1145/3274424}. There have been multiple guides and recommendations for including transgender people in user interface design \cite{morganklausscheuermanHCIGuidelinesGender2020,burtscherWhereWouldEven2020} and in surveys \cite{spielHowBetterGender2019}. These recommendations include allowing all individuals to self-identify their gender, not assuming binary gender, using the language users use, and protecting the privacy of participants. In the case of dealing with medical research and ``embodiment,'' the guidelines recommend measuring physiological parameters such as hormone levels directly, rather than assuming them based on gender.

However, embodiment is a tricky line to draw. When one considers the terms ``sex'' and ``gender,'' the common assumption is that sex is biological and gender is  social. If there is any relationship between the two, it is assumed that sex influences gender, and transgender and intersex people are seen as outliers whose needs vary from ``normal'' populations. However, Springer et al. argue that it is \emph{sex} that cannot be purely decoupled from social factors (i.e. gender) \cite{springerCatalogueDifferencesTheoretical2012}. A ``biosocial turn'' is now beginning in the study of sex and gender \cite{shattuck2019sex}. Many mechanisms that were previously thought to be due to biological ``sex'' differences, are in fact mechanisms that involve differences based on socialization that manifest in biological differences. Springer et al. recommend using ``gender'' to refer to social relations and gender roles and the term ``sex/gender'' to refer to those biological and biosocial factors associated with individual physiology. In this paper, we will use the terms sex/gender and gender, unless we are referring to how these terms are used in a specific work.

\section{Approach}

% add something explicit about standpoint theory or just say it?
Our work draws heavily from transgender studies as an approach, while having some similarity to Black feminist methods, specifically Jennifer Nash's love politics in the form of witnessing \cite{nashBlackFeminismReimagined2019}. We include conversations between the two of us throughout throughout the paper. Personal narrative, especially dialogue, can help uncover ``common pain points and overlooked opportunities'' \cite{ciforGenderedDesignDuoethnographic2020}.  Where duoethnography, used by previous studies, is a research method that employs personal narrative to ``simultaneously generate, interpret, and articulate data'' about a shared experience \cite{norris2008duoethnography}, we include personal narratives throughout this paper to combine, in the words of Susan Stryker, ``the embodied experience of the speaking subject'' (i.e. our experiences using the weight scale) with ``the specialized domain of scholarship'', (i.e. the specifics of the theory and practice of BIA for at home body composition monitoring) \cite{strykerSubjugatedKnowledges2013, spadeResistingMedicineRemodeling2003}. Personal narrative allows for a starting point to a broader conversation about smart weight scales and the implications the system and algorithm design have for technology and biomedical research more broadly.

We are approaching this topic as a White nonbinary person (Kendra), and as a White cisgender woman (Maggie). Neither of us are disabled in ways that are likely to affect our readings from or interactions with the Withings scale. Both of us have considerable background in technology and gender. Kendra is a lawyer teaching at a technology clinic who also teaches in transgender studies. They have, at times, engaged in self-tracking, although not previously around weight. Maggie is an assistant professor at a small liberal arts school where she teaches digital/embedded systems and inclusive engineering design. Her research involves using bioimpedance to help patients manage fluid overload. She is also a self-tracker and has presented her work at several Quantified Self conferences. 

\section{Bioelectrical Impedance Analysis}

Critical analysis of the sort that we deploy in this paper requires the knowledge of how the measurement technology inside smart weight scales works. In this section, we present a background on Bioelectrical Impedance Analysis (BIA). We should note, however, that because the specific testing and equations used by the Withings scale are not publicly available, this background will leverage knowledge from public and peer-reviewed sources and may or may not reflect the specific approaches that the Withings or other consumer-facing scales employ.

At the most basic level, the body can be divided into two main ``compartments:'' fat mass (FM) and fat free mass (FFM) \cite{kyleBioelectricalImpedanceAnalysispart2004,lukaskiEvolutionBioimpedanceCircuitous2013,khalilTheoryFundamentalsBioimpedance2014}. FM includes all the fat in the human body, including what we think of as body fat and also visceral fat around vital organs. FFM is the rest of the tissue; it is about 73\% water, about 7\% bone, and the rest is proteins.

BIA leverages the fact that the water in FFM is conductive; by driving a small, painless current through the body via a pair of electrodes (in weight scales these are two of the electrodes on the scale), the resulting voltage can be measured by another pair of electrodes (also on the scale) and related to the electrical properties of the underlying tissue.

If one assumes the body is a homogeneous cylinder with cross-sectional area A, the measured resistance $R$ (defined as the real part of the measured voltage divided by the current) is equal to:

\begin{equation}
R = \frac{\rho L}{A}    
\end{equation}

where $\rho$ is the conductivity of the cylinder, $L$ is the length of the cylinder, and $A$ is the cross-sectional area of the cylinder. Most BIA equations assume that $L$ is proportional to the body height $H$.  Multiplying both sides of the equation by $L/L$, the resistance can be related to the inverse of the volume, assuming that $V =  L \times A$:
\begin{equation}
R = \frac{\rho L}{A} \cdot \frac{L}{L} = \frac{\rho L^2}{V} 
\end{equation}

If one moves the volume to the other side, there is then:
\begin{equation}
V = \frac{\rho L^2}{R}    
\end{equation}

This volume $V$ corresponds to what is called the ``total body water'' or the volume of all water in the body, which is assumed to be about 73\% of the volume of the FFM. If one multiplies this volume by the presumed density of the FFM to obtain the FFM, the FM and body fat percentage (BF\%) can then be calculated as:

\begin{align}
FM &= Weight - FFM \\
BF\% &= \frac{FM}{FM + FFM} \cdot 100
\end{align}

\subsection{Assumptions of BIA}
\label{sec:BIAAssumptions}

The methods described above require a number of assumptions related to the body. In order for these assumptions to be valid, the resistivity $\rho$ of the measured volume must be homogeneous, and the cross-sectional area must be constant throughout the body such that $V = L \times A$. The assumption that the FFM is 73\% water must also hold, along with the assumed density of the FFM. Finally, it must be assumed that the current penetrates through the whole body in a uniform manner such that the estimated volume is truly reflective of the total body water, and not just a fraction of it.

Of course, these assumptions are not realistic; the body is not a single homogeneous cylinder with precisely known body composition. Instead, BIA leverages different variables that correlate with ``gold standard'' estimations of the FFM to estimate the FFM based on the BIA itself. An example BIA equation for estimating FFM might look like the following:  \cite{kyleSinglePredictionEquation2001}:

\begin{multline}
FFM = -4.104 + (0.518\ \times\  H^2/ R) + (0.231\ \times\ weight) +  (0.130\ \times\ X)\\+ (4.229\ \times\ sex: men = 1, women = 0)        
\label{eq:gendersarebad}
\end{multline}

This equation involves a number of key terms: the $H^2/R$ term, the weight term, the $X$ term (reactance or imaginary part of measured bioimpedance), and sex. Each of these terms is associated with a coefficient in front (along with a constant at the beginning of the equation) that are calculated based on the best fit of the regression equation that minimizes the error between the estimation via BIA and the estimation via the gold standard for the population under test (in this case, the gold standard used was a technology called dual x-ray absorptiometry or DXA). 

Precisely which parameters are included in the regression equations and their corresponding coefficients depends on the population used to calculate the equations and researcher preference. Other researchers also include factors such as the age of the participants and whether or not participants are athletes \cite{khalilTheoryFundamentalsBioimpedance2014}. In some cases these parameters are all incorporated into a single equation (such as the one above that has ``options'' for participant sex), or multiple equations are generated, such as one for ``lean'' participants, one for ``average'' participants, and one for ``obese'' participants \cite{segalLeanBodyMass1988}.

Parameters included in FFM estimation equations often do a lot of ``work'' and their role is not always clearly understood. These parameters and their coefficients ``stand in'' for things such as body density, which can vary depending on both factors included in the equations and those typically excluded from the equations such as ethnicity. For example, age is sometimes included as a parameter because there tends to be a decrease in FFM and an increase in FM with age, and sex is included because males on average have lower FM than females. We unpack these assumptions and coefficient parameters in more depth in Section \ref{sec:critique}.

\section{How BIA Is Marketed}
\textbf{Kendra:} \textit{Of course, I didn't know how BIA worked before using the scale. Nor would looking at the Withings website have revealed any of the fraughtness of BIA to me - when I look at their ads now, they call the technology ``body composition.'' It’s not obvious from their advertising that it’s estimating body fat percentage based on a set of assumptions and an algorithm, rather than providing an individual-level ground truth. If you don’t know how the technology works, it’s actually quite easy to draw the conclusion that the scale just magically knows your actual body fat percentage.}

\textit{Even if I review the ``technical specifications,'' the information contained requires quite a bit of context to determine that what is produced isn't an individualized number. The bullet points say ``Bioelectrical Impedance Analysis / Athlete and non-athlete mode / Unit: body fat \%, total body water \%, muscle mass kg or lb, bone mass kg or lb.'' There’s nothing there that tells me, as an end-user without a lot of expertise in BIA, that it’s engaged in an estimation based on plugging particular values into an equation.} 

\textit{That brings me to the question, Maggie, what were you thinking when you bought the Withings scale? How did the body composition stuff play into it?}

\textbf{Maggie:}  \textit{I've had the scale for a long time - since 2012. That was also the time when the number of people talking about self-tracking was growing, and organizations such as Quantified Self were facilitating some of the first meetups and conferences in this area. Quantified Self emphasized self-knowledge through numbers, often using the frame ``self-research'' or ``personal science'' \cite{wolfConceptualFrameworkPersonal2020}. Over the next few years, the idea of self-tracking would become very hyped, and an entire commercial ecosystem Whitney Erin Boesel dubs ``quantified self'' (i.e. little q, little s, vs big Q, big S) was formed \cite{boeselWhatQuantifiedSelf2013}. Looking back at my data, my first weigh in was March 23rd, 2012. I wanted to learn more about the tech that was out there and see what I could do to make sense of things, and was also inspired by a Quantified Self ``Show \& Tell'' talk by Amelia Greenhall about how she weighed herself everyday and sustained her weight loss long term \cite{greenhallWeightWeightDon2013}. I was excited about the possibility of self-tracking and consistent habits improving my fitness and my life. I wanted to learn more and then translate that knowledge to help others.}

\textbf{Kendra:} \textit{This idea of self-knowledge is really exciting. That’s what I was hoping for when I stepped on the scale as well - some numbers to help me quantify what I was feeling about my body. But of course, that’s not what I got. As a White nonbinary person, what I learned is that this tech isn't build for me - in part because of the choices that technology companies make, and in part because of the failure to meaningfully account for transgender experience as part of the underlying clinical testing. And it’s worse for non-White nonbinary or intersex folks, who are both not represented in the studies in terms of sex/gender or race/ethnicity. So much for smart scales.}

\section{Critiques of BIA}
\label{sec:critique}

Many limitations of BIA have been well established in the medical literature \cite{khalilTheoryFundamentalsBioimpedance2014}, though some researchers argue that the techniques are still appropriate to use under the correct circumstances \cite{wardBioelectricalImpedanceAnalysis2019}. Researchers suggest caution when using BIA, especially when working with populations that have altered body composition, such as patients with fluid overload. In these cases, some researchers have developed equations specifically for a particular patient population, or have used alternative methods of body composition assessment that don't rely on regressions (see e.g. \cite{keaneBodyCompositionMonitor2017}).

A major challenge with body composition using BIA is that the two compartment model of ``fat mass'' (FM) vs ``fat free mass'' (FFM) inherently requires assumptions about the composition of the two different compartments (in addition to other assumptions such as homogeneous composition as discussed previously in Section \ref{sec:BIAAssumptions}). Uniformity of the FM is a fairly reasonable assumption across individuals \cite{martinAdiposeTissueDensity1994}, but assumptions about the composition of the FFM are not \cite{coteEffectBoneDensity1993}. Additionally, when using the regression equations, the assumptions about the composition of the FFM are obscured into the coefficients associated with the included variables such as age, weight, and sex. 

Because the linear regressions are optimized at a population level and most studies do not examine the accuracy of the estimations at an individual level, there is no guarantee that a specific equation is accurate for any given individual. Additionally, once one begins to consider any individual that is not perfectly matched for the population that was used to create the equations, the role of these variables becomes increasingly murky but also increasingly important in order to design equations that work for populations historically not included in the populations used to generate the equations. This includes non-White people, trans and nonbinary people, intersex people, people with chronic illnesses, and those at the intersections of these categories.

\subsection{Unpacking ``Sex''}

We begin with the assumptions and role of ``sex'' in the equations.\footnote{We use the term ``sex'' here given that this is the term used by the researchers. However, sex and gender are entangled as described in Section \ref{sec:background}.} Sex in BIA is either coded as 0 for female and 1 for male (effectively changing the offset of the equations, as in Equation \ref{eq:gendersarebad}) or there are separate equations created for male participants and female participants, as in Sun et al. \cite{sunDevelopmentBioelectricalImpedance2003a}. What the literature means by ``male'' or ``female'' is unclear, and these terms are often confounded with gender identities of ``man'' and ``woman'' as in Equation \ref{eq:gendersarebad}. As we discussed in Section \ref{sec:background}, ``sex'' as a concept is just as fraught and contingent as gender \cite{fausto-sterlingSexingBodyGender2000,springerCatalogueDifferencesTheoretical2012,davisIntersexSocialConstruction2017a}.  This is not a problem unique to (or caused by) the existence of trans (or intersex) people. 

Although the methods by which ``sex'' was evaluated in the BIA literature is unclear, it is common for reported sex to be a participant's sex assigned at birth. And ``sex assigned at birth'' is generally only a proxy for someone’s external genitalia at birth, which is only one of the many characteristics that are often encompassed under the word sex \cite{davisIntersexSocialConstruction2017a,fausto-sterlingSexingBodyGender2000}. Others include hormone balance, chromosomal makeup, and internal reproductive organs. We could not find an example of a study that produces BIA estimates that discusses what sexual characteristics they round up to a determination of sex, and it is generally not clear how the identification of sex was made (i.e. whether self-identified or identified by the researchers). This lack of specificity is one of the first and most significant barriers to creating a more inclusive algorithm for transgender people. Given how large the sample sizes were for some of the populations used to create BIA equations (upwards of 5,000 in  \cite{kyleFatfreeFatMass2001}), it is statistically unlikely that no transgender people were involved. But we don’t know how they were accounted for or counted in the calculations.

The use of ``sex'' itself is also a ``stand in'' for other parameters. What the researchers presumably intend is for ``sex'' to stand in for the roughly dimorphic variation in body composition between those assigned male at birth and those assigned female at birth  \cite{davisIntersexSocialConstruction2017a}. However, it is not the fact that a person is assigned male at birth (i.e., they have genital erectile tissue of a certain size) that makes those assigned male at birth have lower FM on average compared with those assigned female at birth. In fact, FM and the distribution of fat may actually be biosocial phenomena: they may depend on both sex AND gender (sex/gender). For example, one effect of larger quantities of testosterone is changes in fat distribution and increases in muscle production, resulting in lower body fat percentages \cite{spanosEffectsGenderaffirmingHormone2020}. However, most research into sex differences in body composition do not explore other explanations, such as the differences in diet between men and women in some cultural contexts \cite{prattala2007gender,holm2000role}. The BIA literature also does not disaggregate sociocultural context that might be caught up in the word ``sex'', nor account for the myriad ways in which someone assigned male at birth might not match the presumed physiological composition associated with those assigned male at birth on average. 

With only two sex or gender options, some intersex and most nonbinary users experience what D.E. Wittkower has termed a dysaffordance  -- they have to pretend to be something they are not in order to access the service \cite{wittkowerPrinciplesAntidiscriminatoryDesign2016,costanza-chockDesignJusticeCommunityLed2020}. The lack of transparency in the BIA equations makes it difficult to tell exactly what adjustments might need to be made to make equations more inclusive, or to at the least advise individuals as to which ``sex'' they should choose when using consumer technologies that incorporate BIA. On the other hand, the current setup of the Withings scale that asks for user gender may actually produce more accurate results for trans women and trans men who have undergone gender-affirming hormone therapy, as their body composition may resemble that of a cis woman or cis man, rather than their sex assigned at birth \cite{klaverChangesRegionalBody2018,spanosEffectsGenderaffirmingHormone2020}.

\subsection{Other Issues With Published Studies}
\label{sec:otherIssues}

White nonbinary or intersex people, despite experiencing a dysaffordance, may be better off than their siblings of color because the composition of the study populations used in BIA equations in the Withings scale are unclear, and there is no way to enter racial or ethnic information.\footnote{We use both the terms race and ethnicity because the literature is mixed on which are relevant factors, likely because of the entanglement of biological and social factors similar to sex/gender as discussed in Sections \ref{sec:background} and  \ref{sec:critique}.} The majority of measurement studies involving BIA include primarily Caucasian subjects and assume that Caucasian subjects are to serve as a ``reference'' for other ethnicities.\footnote{The use of White people as a default mirrors Kimberle Crenshaw's critiques of US law, as well as Ruha Benjamin's critique of photographic film \cite{benjaminRaceTechnologyAbolitionist2019,crenshawDemarginalizingIntersectionRace1989}.} This is an issue because body composition can vary among ethnic and racial groups due to the environment, diet, cultural factors, and anthropomorphic measurements such as limb length and body size \cite{deurenbergBodyMassIndex1991}. Researchers or device manufacturers interested in using BIA equations validated in a Caucasian population therefore need to cross-validate the equations in the population of study. 
As with sex, race or ethnicity is likely standing in as a proxy for some other variables, such as environmental racism, socioeconomic status, or some actual relationship between FFM composition and certain genetic factors. Some studies suggest that ethnicity specific compensations are needed to use body composition equations in different ethnic groups \cite{cohnBodyElementalComposition1977,schutteDensityLeanBody1984}, whereas other studies have shown that stratification based on adiposity (i.e. body fat percentage) is more important than race \cite{ainsworthPredictiveAccuracyBioimpedance1997}. Regardless, cross-validation studies are necessary to ensure that the assumptions of equations produce accurate results for all populations.

Another major issue with BIA is that the regression equations themselves are validated against ``gold standards'' that in turn have their own assumptions. For example, BIA equations are often compared against air displacement plethysmography or hydrostatic weighing. But both of  these techniques require assumptions about the density of the FFM. The density of the FFM depends on numerous factors such as age, sex, and ethnicity \cite{ellisHumanBodyComposition2000}. Therefore any ``gold standards'' that likewise depend on the two-compartment model (i.e. FM and FFM) are themselves are only valid in the same population, and most of those are also primarily tested on White (presumably cis and binary gender) subjects.

Even techniques that do not depend on the two-compartment model such as dual x-ray absorptiometry (DXA) are found to significantly underestimate fat mass when compared with CT scans \cite{kullbergWholebodyAdiposeTissue2009}. Ultimately, the only way to validate a body composition device is using cadaver studies and chemical analysis, which have some of the same issues - the results would then only be validated for those similar to the cadavers \cite{shepherdBodyCompositionDXA2017}. Although measurement problems of this type are common in all areas of science, the lack of physiological basic science about how and why FFM varies with respect to both social and biological factors in an intersectional way means that it is difficult to determine which assumptions will hold across populations. Because of this, the lack of population-specific testing in each individual ``gold standard'' testing regime complicates the possibility of meaningful validation for folks who do not fit the default.

\subsection{Why (anti-)Black Boxes? The Lack of Regulatory Oversight}

The gap between promises and reality in the smart medical device space is in part due to the limited scope of review by the United States' Food and Drug Administration (FDA). The vast majority of medical devices, including ``smart'' and ``connected'' devices, are approved through a program at the FDA known as the 510k approval process. This process requires no clinical trials and very little oversight - device manufacturers merely need to prove that their product is ``substantially equivalent'' to that of an already approved medical device. The Withings scale discussed in this paper received 510(k) clearance \cite{Withings510Summary2012}; similar scales on the market such as the Fitbit Aria scale received approval even after the devices were already on the market \cite{FitbitAria5102014}. In 2014, the FDA announced their intentions to cease requiring even 510(k) approval for devices such as smart weight scales. This means that there is very little regulation of these devices, and certainly no required clinical validation of the algorithms used to calculate body composition, despite the label of ``clinically tested'' on the Withings website.

This lack of regulatory oversight results in most consumer-focused deployments of technologies like BIA being ``black box'' algorithms. Popularized in the context of technology studies by scholar Bruno Latour, an algorithm is considered a black box ``when a statement is simply presented as raw fact without any reference to its genesis or even its author.'' \cite{harman2010prince}. When using the Withings scale, only the final body fat percentage is made available, and there is no explicit reference to an algorithm in the app or in most of the marketing materials. Additionally, Withings does not release the equations its scales use to calculate FFM or the populations that it used to calculate those equations. The power of the black box is that we cannot thoroughly investigate a subject about which nothing is known \cite{pasqualeBlackBoxSociety2015}. We are left with the assumption that, unless proven otherwise, Withings' internal algorithm production mirrors the biases of the research at large, but again, it is impossible to tell. All that we know are the inputs that Withings asks users for (binary gender, height, age and athlete status).

We can draw some conclusions even with just publicly available information. The binary approach to the gender question without any explanatory text both erases nonbinary people, and does not ask the right question about the embodiment of the user. The Withings scale does not prompt the user to enter information related to ethnicity, so it is not possible that the scale is using an equation that adjusts variables to compensate for different FFM factors in racial or ethnic groups. Because of the marketing of the technology, non-White users may not even know that it might be relevant. The Withings scale’s algorithm is, in the words of Ruha Benjamin, an anti-Black box \cite{benjaminRaceTechnologyAbolitionist2019}. 

Furthermore, what evidence we do have points to BIA being unreliable even on the populations that it is theoretically well positioned for. We reviewed the list of studies that Withings shared on their page for researchers and found only one study that specifically evaluated the body composition aspects of the scales \cite{collierWithingsBodyCardio2020}. The results of the study compared the performance of the newer Body Cardio scales with an air displacement plethysmography device called the Bod Pod (a ``gold standard'' discussed in Section \ref{sec:otherIssues}). The mean absolute percentage error of the body fat estimation of the Body Cardio weight scale compared with the Bod Pod was greater than 25\%, well above a previously established 1.5\% threshold deemed as an acceptable error \cite{collierWithingsBodyCardio2020}.\footnote{It is important to note, however, that while air displacement plethysmography is often used as a comparison device/gold standard, it relies on assumptions that can compromise its effectiveness as we discuss in Section \ref{sec:BIAAssumptions}. Withings also argues that these scales should be used to indicate trends rather than for absolute assessment \cite{worleyComparingAccuracyBody2016}.} This suggests that any results from the Withings scales should be interpreted with extreme caution, even on the target population who is most well represented in the studies likely used to create equations.  

\section{Denying Self-Knowledge}

\textbf{Kendra:} \textit{The distance between Withings’ promise of self-knowledge and the reality of regression equations is upsetting. They advertise all of these benefits to self-quantification, but it’s actually limited by the technology \cite{felberReasonsQuantifyYourself2012}. As with many algorithmic technologies, the creation of regression equations based on a limited sample cannot and will not create accurate self-knowledge amongst those who do not fit within those samples.} 

\textit{If the scale was actually individually calculating a ground truth number, as opposed to using a regression based on height and skirt-wearing vs. pants-wearing, being nonbinary wouldn't matter! To be honest, 90\% of the time when I’m asked my gender or sex, it doesn't matter. It’s an arbitrary box checking exercise. So why would the Withings scale be different?}

\textit{There’s this sleight of hand involved in not revealing to people how the technology works that creates the situation in which a nonbinary person could go in expecting self-knowledge and getting a prediction totally disconnected from what they themselves could tell you about their body. I could have told the scale more about me to make the answer more accurate, but that wasn't an option. I don’t necessarily mind sharing information about my hormone status, my sex assigned at birth, or other facts about my body if they’re actually useful. And although my views on volunteering race are shaped by my membership in a privileged racial group, I suspect many users would prefer to share race or ethnicity information if it meant that they would get more accurate results.}

\textit{Withings asks for my gender, but it doesn't want it, even aside from the app’s confusion between gender and sex. I know things about myself that are relevant to its guessing, but there’s no way to translate this knowledge within the limited frame of reference produced by the clinical trials. There’s no way to come out with a more accurate picture or contest the bounded understanding of the system. That feels erasing, even more than the mere existence of the binary prompt.}

\textit{It makes me wonder about all of the other random gender/sex requests that I've encountered when using technologies around health. Does Fitbit want my gender/sex for calorie burning calculation? Does Apple? What about the Ring Fit? How deep does the rabbit hole go?}

\section{Paths Forward and Recommendations}

Trans and nonbinary people of all races deserve to have access to inclusive self-tracking technologies that do not collapse their identities or force them to ignore relevant self-knowledge. What can be done to improve these technologies? We evaluate three options for how to handle sex/gender under the hood of a BIA-calculating device such as the Withings scale, and then provide overall recommendations as to how to handle the use of regression equations based on limited medical testing.

\begin{figure}[h]
  \centering
  \includegraphics[width=0.9\linewidth]{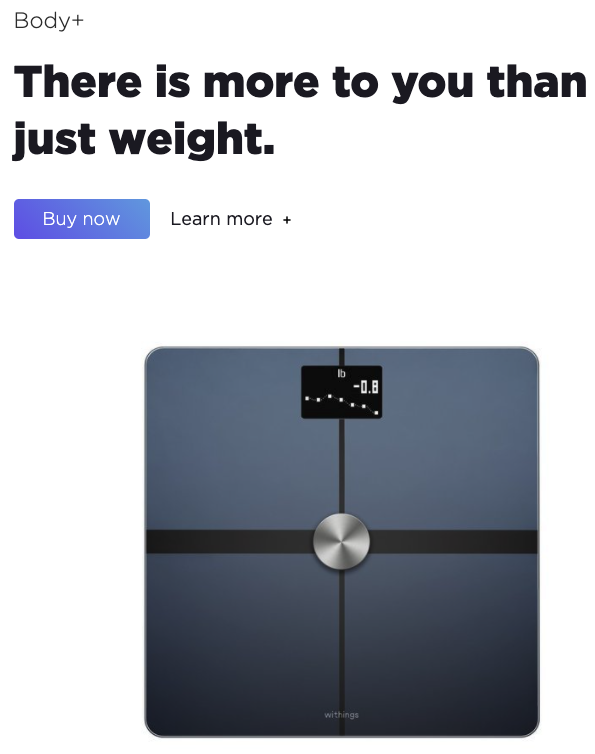}
  \caption{A portion of the front page of the Withings website advertising a version of their smart scale.}
  \Description{Image of the front page of the Withings website advertising a square weight scale. The text reads ``There is more to you than just weight.''}
  \label{fig:morethanweight}
\end{figure}

It would be inappropriate to continue without noting that BIA as a technology deployed in smart scales may be fundamentally inseparable from the fatphobic medical and cultural context that has created concerns about body fat in the first place \cite{lupton2018fat}. Withings may claim that ``there is more to you than just weight'' (see Figure \ref{fig:morethanweight}), but the subtext of its advertising indicates that you should want to weigh less. That is not fixable through recommendations around the handling of sex (or gender, or hormone status). It might be reasonably asked - given all its flaws, should BIA be used in consumer devices at all? We don’t seek to answer that question in this paper. Our aims are more modest. Nonbinary folk and transgender folks deserve access to technologies of self-knowledge, even as those technologies may be used both by companies and individuals to suggest harmful body standards.

We provide a series of options for making weight scales based on BIA more inclusive, with recommendations for users, and both manufacturers and researchers. Most of our recommendations specifically focus on sex/gender, but it is worth noting that overall, BIA also has a long way to go when it comes to race/ethnicity, which we leave for future work to explore in more depth. Although our recommendations are designed based on the specific context of BIA smart scales, they could potentially apply to many areas where binary sex/gender is used as part of an algorithmic system.

\subsection{Option 1: Eliminate Sex/Gender as A Variable}

One option for making systems more inclusive of nonbinary people would be the elimination of sex/gender as a variable, using one equation for all users. Practically speaking, this is not difficult. Manufacturers would not have to acquire new data, only re-run the regression to find the best fit without sex/gender as a variable in the regression equation(s). 

However, a drawback of this option is that elimination of sex / gender as a variable for all users would result in readings that were less accurate in aggregate, as the inclusion of sex does reduce the mean error of the regression equations \cite{khalilTheoryFundamentalsBioimpedance2014}. Given the lack of accuracy of the technology as a whole, this may or may not be hugely significant - however, users who find a sex/gender binary appropriate for their bodies might be upset to lose accuracy in order to make the technology more inclusive.

\subsection{Option 2: Add a Third Option}

The next option is to add a third option to the menu where one currently chooses sex/gender. One method of implementing a nonbinary option would be the optional elimination of sex/gender as a variable as listed above for people who select a third option. There would then be three options: ``male,'' ``female,'' and ``sex/gender neutral.'' 

This option could be helpful for some intersex people, nonbinary people who have not medically transitioned but who would prefer potentially less accurate results to having to volunteer a binary marker, nonbinary people who have taken some medical transition steps, and anyone else for whom the binary sex options are unlikely to produce accurate results. 
A cautionary note: having a third option in the menu does not a nonbinary inclusive app make.  As Anna Lauren Hoffman explains in her generative work on data violence and inclusion, without taking meaningful steps to change the power dynamics present in the product, inclusion is at best, lip service \cite{hoffmannTermsInclusionData2020}. For example, when Facebook allowed for self-identification with a wider variety of gender identities on their platform, they did not fundamentally change the binary advertising logic of male or female, making their claims of nonbinary inclusion questionable \cite{bivensGenderBinaryWill2017}. Thus, adding a third option is only appropriate if there is an underlying algorithmic change.

\subsection{Option 3: Stop Using Sex/Gender as a Proxy}

Ultimately, the ideal outcome of this work would be for the field to take a step back and consider the role that sex/gender are playing as a ``stand in'' for things like body fat distribution and anthropomorphic information. This is exactly the kind of work that HCI researchers have recommended when considering trans embodiment \cite{morganklausscheuermanHCIGuidelinesGender2020,burtscherWhereWouldEven2020}. But it is more difficult in this case than many others because of how pervasive assumptions about sex and gender are in clinical research \cite{tannenbaumSexGenderAnalysis2019,springerCatalogueDifferencesTheoretical2012,mosesonImperativeTransgenderGender2020}. The full, complicated role that sex and gender play in BIA equations and beyond are not well understood. Significant fundamental research is necessary to begin to understand which additional factors to measure and how to measure them in cost-effective and reliable ways.

A deeper understanding of sex/gender and body composition will require ``slow science'' \cite{stengers2018another}. With more information about the role that these factors are playing, additional information could be provided by end users - everything from body shape (i.e., ``apple'', ``pear'') to hormone status. This information could even be made optional to not place an additional burden on those unfamiliar with the specifics or who want to do the basics. Fundamentally, this approach is the most well aligned with the promises that companies such as Withings make about their technologies, but would also require the most fundamental research.

\begin{table*}[htbp]
  \centering
  \caption{Recommended best practices for trans, nonbinary and intersex inclusion in regression based technologies such as BIA. Table design inspired by \cite{mosesonImperativeTransgenderGender2020}.}
    \begin{tabular}{rrp{29em}}
    \toprule
    \multicolumn{1}{c}{\textbf{Context}} & \multicolumn{1}{c}{\textbf{Marginalizing Practices}} & \multicolumn{1}{c}{\textbf{Inclusive Practices}} \\
    \midrule
        \multicolumn{1}{p{7.585em}}{For manufacturers and researchers} & \multicolumn{1}{p{15.75em}}{Assume sex is purely biological and gender is purely social.} & Review literature on sex/gender and select the appropriate measures for the specific project. If none are available, consider conducting more basic research, and consider and articulate the limitations of the state of the art. \\
          &       & \multicolumn{1}{r}{} \\
          &       & Assume a biosocial explanation for physiological differences unless evidence clearly suggests otherwise. \\
          &       & \multicolumn{1}{r}{} \\
          & \multicolumn{1}{p{15.75em}}{Ignore the existence of trans, non-binary, and intersex people.} & Acknowledge the existence of trans, non-binary, and intersex people and how their physiology or experiences might be different from other users. \\
          &       & \multicolumn{1}{r}{} \\
          &       & Acknowledge, if needed, the limitations of the current results and how trans and nonbinary people can still obtain the best results for them. \\
          &       & \multicolumn{1}{r}{} \\
          & \multicolumn{1}{p{15.75em}}{Analyze results only at the population level.} & Analyze results for sub-groups and at the individual level. \\
          &       & \multicolumn{1}{r}{} \\
          &       & Evaluate results across racial and ethnic groups, implementing race/ethnicity selection or inclusion of non-proxy variable as appropriate. \\
          &       & \multicolumn{1}{r}{} \\
    \multicolumn{1}{p{7.585em}}{For manufacturers} & \multicolumn{1}{p{15.75em}}{Elide the measurement precision and assumptions.} & Explicitly state the precision of the measurement system, along with assumptions and constants used. \\
          &       & \multicolumn{1}{r}{} \\
          & \multicolumn{1}{p{15.75em}}{Represent sex and/or gender with pictograms.} & Use clear terminology based on the underlying research or known physiology to select a term. \\
          &       & \multicolumn{1}{r}{} \\
          &       & Be explicit why sex and/or gender are being used so that trans, non-binary and/or intersex users can choose the best option for them. \\
          &       & \multicolumn{1}{r}{} \\
    \multicolumn{1}{p{7.585em}}{For researchers} & \multicolumn{1}{p{15.75em}}{Assume ``gold standard'' has no built-in assumptions.} & Discuss the assumptions embedded in the gold standard methods themselves, and how those assumptions influence the results. \\
          &       & \multicolumn{1}{r}{} \\
          & \multicolumn{1}{p{15.75em}}{Assume the collection of sex and/or gender information is obvious or straightforward, and therefore does not need to be discussed in depth.} & Include whether reported sex and/or gender was based on self-report, and if so, what options were available for participants to choose between. Were there only two options? Was there a free response option? \\
          &       & \multicolumn{1}{r}{} \\
          &       & Explicitly state how sex and gender are being used and what they stand in for. Example: ``sex is a stand-in for the dimorphic distribution of body fat in the human population.'' \\
          \bottomrule
    \end{tabular}%
  \label{tab:recommmendations}%
\end{table*}%

\subsection{Recommendations}

First, some advice to transgender, nonbinary, and intersex people who wish to use technology that incorporates BIA but presents binary sex options. Based on studies looking at changes in body composition, hormone status is a very significant variable for body composition of the bodily characteristics, perhaps more significant than other variables that are encompassed by the word sex \cite{spanosEffectsGenderaffirmingHormone2020,elbersReversalSexDifference1997}. So we would recommend that if folks must pick a binary category, they pick the one most closely aligned with their hormonal balance - male if in a normal male range for testosterone, female if not. In any case, because of the study populations used to produce equations and the black box nature of these algorithms, the actual value produced is unlikely to be accurate and should be used primarily to track change over time rather than for its absolute value. (This recommendation also holds true for any person of any gender using the scale.)

In Table \ref{tab:recommmendations}, we lay out additional recommendations for researchers and manufacturers who wish to build more inclusive regression-based technologies. Elaboration on some of these recommendations and additional recommendations for different contexts can be found in the following references: \cite{tannenbaumSexGenderAnalysis2019,mosesonImperativeTransgenderGender2020,morganklausscheuermanHCIGuidelinesGender2020,springerCatalogueDifferencesTheoretical2012,spielHowBetterGender2019,burtscherWhereWouldEven2020,gebruOxfordHandbookAI2019}. 

In general our recommendations can be summarized as a) acknowledge the existence of non-gender normative people, b) make fewer assumptions, and c) explain in more detail the limitations of technology. Of course, it may be difficult for companies to fully be honest about measurement accuracy and precision. But if being explicit and honest with customers about these errors and assumptions would make them think twice about purchasing a product, perhaps the best next step is for companies to reconsider their business model.

Of course, including trans people after the fact is not ideal. Participatory methods that incorporate transgender people in problem definition around medical devices - to design, in the words of Haimson et al., trans technologies - would be preferable to all of these stopgap measures \cite{haimsonDesigningTransTechnology2020,costanza-chockDesignJusticeCommunityLed2020,normanshamasLookingMarginsIncorporating2019,gebruOxfordHandbookAI2019}. However, as practitioners who work with participatory methods, we understand that such practices are unlikely to arise overnight. Until it's widely accepted that designing for those on the margins can create better medical devices, participatory design may never fully adopted by those who commercialize them. 

\section{Conclusion}
It can be easy to assume that the use of ``sex'' in quasi-medical applications is neutral, just another fact about one’s body that allows for a more accurate complete picture. But, in the immortal words of dril, ``this whole thing smacks of gender'' \cite{drilThisWholeThing2012}. When the lived realities of nonbinary folks cause us to scratch below the surface, the lack of careful thought around assumptions that go into technologies like smart scales becomes clear. Cultural beliefs about gender are driving the bus when it comes to engagement with ``sex differences.'' And because of that, sex, even in clinically tested BIA equations, is holding space for too many variables, supported by too little basic research. 

When inadequately validated, (anti)-Black box algorithms built on these shaky foundations deceive their users. They harm people who do not line up with the assumed user base, promising knowledge of self but instead merely reproducing the violence of erasing clinical systems \cite{hoffmannTermsInclusionData2020}. 

It doesn't have to be this way. Even without additional clinical testing or regulation, there are clear steps that manufacturers can take to mitigate some of the harms caused by these systems.  First, they can educate users as to the population-level accuracy of metrics like BIA, rather than advertising body composition analysis as if it was accurate on an individual basis. Second, as discussed above, they can make clear how the technology does or does not work on transgender or nonbinary people, while also identifying other factors (such as race and/or ethnicity) that make results more or less accurate.  Finally, manufacturers could release as much information about their equations as possible, including validation studies, in order to facilitate cross-validation by independent researchers. 

Admittedly, these solutions may reify technical expertise and serve to legitimize the ideas of these types of body measurement, as Greene, Stark, and Hoffman, point out in their work on technological ethics \cite{greeneBetterNicerClearer2019}. Ultimately, BIA is just one example of how regression-based body measurements, whether implemented in technologies like smart scales or described in scientific papers, harm those who are not presumed to be the ideal. And that whole thing is worth overturning.

\begin{acks}
Thank you to Siobhan Kelly for their perceptive comments on lip service, and to the many essential workers, particularly those at Philly Foodworks and South Square Market, whose labor allowed us to write (and eat).
\end{acks}

%\clearpage
\bibliographystyle{ACM-Reference-Format}
\bibliography{sample-base}

\end{document}